\documentclass[aip,jap,twocolumn,reprint,preprintnumbers,amsmath,amssymb]{revtex4-1}

\usepackage[dvips]{graphicx}
\usepackage{dcolumn}
\usepackage{bm}

\begin{document}

\preprint{Journal ref.: J.\ Appl.\ Phys.\ \textbf{113}, 233702 (2013)}

\title{Empirical relationship between x-ray photoemission spectra and electrical conductivity in a colossal magnetoresistive manganite La$_{1-x}$Sr$_x$MnO$_3$}

\author{T.~Hishida}
\email[] {t-hishida@mg.ngkntk.co.jp}
\author{K.~Ohbayashi}
\affiliation{NGK SPARK PLUG CO., LTD., 2808 Iwasaki, Komaki, Aichi 485-8510, Japan}

\author{M.~Kobata}
\affiliation{National Institute for Materials Science (NIMS), 1-1-1 Kouto, Sayo, Hyogo 679-5198, Japan}
\affiliation{Condenced Matter Science Division, Japan Atomic Energy Agency, 1-1-1 Kouto, Sayo, Hyogo 679-5148, Japan}

\author{E.~Ikenaga}
\author{T.~Sugiyama}
\affiliation{Japan Synchrotron Radiation Research Institute (JASRI), 1-1-1 Kouto, Sayo, Hyogo 679-5198, Japan}

\author{K.~Kobayashi}
\affiliation{Condenced Matter Science Division, Japan Atomic Energy Agency, 1-1-1 Kouto, Sayo, Hyogo 679-5148, Japan}
\affiliation{Hiroshima Synchrotron Radiation Center, Hiroshima University, 2-313 Kagamiyama, Higashi-Hiroshima, Hiroshima 739-0046, Japan}

\author{M.~Okawa}
\author{T.~Saitoh}
\email[] {t-saitoh@rs.kagu.tus.ac.jp}
\affiliation{Department of Applied Physics, Tokyo University of Science, 6-3-1 Niijuku, Katsushika, Tokyo 125-8585, Japan}

\date{\today}

\begin{abstract}
By using laboratory x-ray photoemission spectroscopy (XPS) and hard x-ray photoemission spectroscopy (HX-PES) at a synchrotron facility, we report an empirical semi-quantitative relationship between the valence/core-level x-ray photoemission spectral weight and electrical conductivity in La$_{1-x}$Sr$_x$MnO$_3$ as a function of $x$. 
	In the Mn 2$p_{3/2}$ HX-PES spectra, we observed the shoulder structure due to the Mn$^{3+}$ well-screened state. However, the intensity at $x$=0.8 was too small to explain its higher electrical conductivity than $x$=0.0, which confirms our recent analysis on the Mn 2$p_{3/2}$ XPS spectra.
	The near-Fermi level XPS spectral weight was found to be a measure of the variation of electrical conductivity with $x$ in spite of a far lower energy resolution compared with the energy scale of the quasiparticle (coherent) peak because of the concurrent change of the coherent and incoherent spectral weight.


\end{abstract}



\maketitle

\section{INTRODUCTION}

	The perovskite oxide La$_{1-x}$Sr$_{x}$MnO$_{3}$ (LSMO) is a prototype of colossal magnetoresistance (CMR) effects in the family of Mn oxides $(R,A)$MnO$_{3}$ or $(R,A)_3$Mn$_2$O$_7$ ($R$=rare earths, $A$=alkaline earths)
and thus have been intensively studied
for over two decades.\cite{vonHelmolt93,Tokura94,Cohn97,Averitt01,Moritomo96,Chainani93,Saitoh95}
	The remarkable phenomenon CMR, characteristic of LSMO, is directly connected with its unique electronic structure, namely the double exchange (DE) mechanism.\cite{Zener51} However, it is also well-known that DE alone cannot explain \textit{colossal} MR effects\cite{Millis95} 
though the missing components are
still under debate. Hence, further electronic-structure studies are required for a complete understanding of the CMR effects as well as further development of magnetoresistive materials in device applications.

	Photoemission spectroscopy is one of the most direct probes that can be used to investigate the electronic structure of materials.
	A recent remarkable discovery in this field is the ``metallic" peak/shoulder (hereafter shoulder, for simplicity) in the Mn 2$p_{3/2}$ core-level spectra of LSMO thin films ($0\leq x\leq 0.55$) measured by hard x-ray photoemission spectroscopy (HX-PES).\cite{Horiba04} This shoulder structure is interpreted as the well-screened peak of the Mn$^{3+}$ state and its intensity is considered to represent 
the metallicity of the system.\cite{Horiba04}
	The shoulder intensity was even found to scale the conductivity in La$_{0.85}$Ba$_{0.15}$MnO$_3$ thin films.\cite{Ueda09} However, the full $x$ dependence of the shoulder intensity and its relation to electrical conductivity was not resolved.
	Recently, Hishida \textit{et al.}\cite{Hishida12} found that this shoulder is observable using laboratory x-ray photoemission (XPS) using an Al $K\alpha$ x-ray source.
	They investigated the shoulder structure in the full range of $x$ and found that the $x$-dependence of the shoulder intensity did not follow the $x$-dependence of  the electrical conductivity. This is because the shoulder due to the Mn$^{3+}$ state loses its weight in a larger $x$, while the electrical conductivity does not drop in proportion to $x$.\cite{Urushibara95}
	Employing a new multiple-peak fitting strategy, they extracted a Mn$^{4+}$ well-screened peak hidden behind the Mn$^{3+}$ main peak and found that the electrical conductivity was represented by the sum of the spectral weight of both the Mn$^{3+}$ and Mn$^{4+}$ well-screened states. However, the multiple-peak fitting for a large $x$ was not perfect because of the very small Mn$^{3+}$ well-screened shoulder. Hence a further investigation was still needed.


	One may expect that the electronic states responsible for electrical conductivity are completely reflected in the quasiparticle spectral weight in the vicinity of the Fermi level ($E_{\rm F}$). However, this is not always true, particularly in strongly correlated electron systems like LSMO.
	Although the low-temperature electrical conductivity for LSMO is as high as about $10^{5}$ ($\Omega$ cm)$^{-1}$ at $x = 0.4$,\cite{Urushibara95} angle-integrated photoemission studies have failed to observe large near-$E_{\rm{F}}$ spectral weight\cite{Sarma96,Park96} as can be seen in Ti or V oxides.\cite{Yoshida02}
Instead, only a few angle-resolved studies on layered manganites have succeeded in observing an extremely small quasiparticle peak in very limited $k$-space areas.\cite{Mannella05,Sun06}
	Nevertheless, the temperature dependence of the near-$E_{\rm{F}}$ spectral weight over hundreds meV was found to follow that of the Drude weight or electrical conductivity,\cite{Saitoh00} reproducing the agreement between the temperature dependence of the quasiparticle peak weight (with the width of $\sim$50 meV) and that of the electrical conductivity.\cite{Mannella07} Again, there have been no $x$-dependent studies on this issue so far.

	In this research, we investigate the electronic structure of La$_{1-x}$Sr$_x$MnO$_3$ by XPS and HX-PES in order to determine a relationship between near-$E_{\rm F}$/Mn 2$p_{3/2}$ core-level XPS spectral weight and the electrical conductivity in the full range of $x$.


\section{Experimental}

Polycrystalline samples of La$_{1-x}$Sr$_x$MnO$_3$ ($x$=0.0, 0.1, 0.2, 0.33, 0.4, 0.5, 0.55, 0.67, 0.8, 0.9, and 1.0) were prepared by solid-state reaction. Starting powders of La(OH)$_3$,
SrCO$_3$,
and Mn$_2$O$_3$ 
were weighed in specific proportions and then ball-milled for 15h using zirconia balls and ethanol. The mixed powders were dried and heated at 1100$^{\circ}$C for 2h in air. The solid masses obtained after cooling to room temperature were crushed in the ball milling again for 15h. The powders were dried and pressed into pellet under an isostatic pressure of 0.8 ton/cm$^2$, and then sintered at 1600$^{\circ}$C for 1h in air. 

	The HX-PES experiments were carried out at the undulator beam line BL47XU in SPring-8 using a VG-Scienta R4000  analyzer with a wide acceptance objective lens\cite{Kobata10} and an excitation energy of 7939.9 eV with a bandwidth of 0.29 eV. The XPS measurements were performed with a PHI Quantera SXM instrument (base pressure $5\times 10^{-9}$ Torr) using a monochromatic Al $K\alpha$ source ($h\nu=1486.6$ eV) with the total energy resolution of about 0.64 eV in FWHM. The analyzer pass energy was set to 55 eV for narrow scans. In order to obtain fresh, clean surfaces, the samples were fractured in the prep chamber at room temperature immediately before the measurements. The binding energy was corrected by using the value of 84.0 eV for the Au 4$f_{7/2}$ core-level spectrum. The measurement vacuum was better than $1\times 10^{-8}$ Torr.
	No detectable C 1$s$ peak was observed from the prepared surfaces and the quality of the O 1$s$ peak was comparable or better than literature.\cite{Bindu11,Suppl}

\section{Results and discussion}


\begin{figure}[t]
	\begin{center}
 	\includegraphics[width=80mm,keepaspectratio]{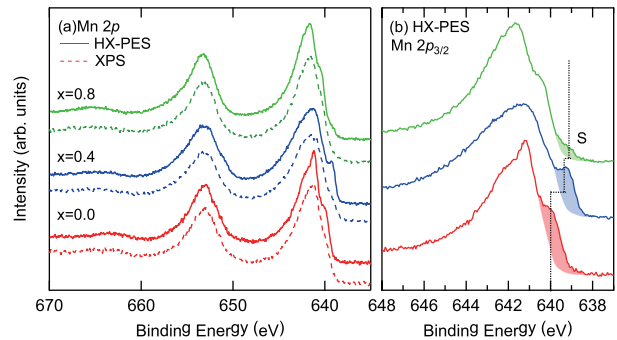}
  \caption{(a) Mn 2$p$ core-level HX-PES and XPS spectra of LSMO of $x = 0.0$, 0.4, and 0.8. (b) HX-PES spectra in the Mn 2$p_{3/2}$ region. S denotes the Mn$^{3+}$ well-screened state.}
\label{MnHXPES}
\end{center}
\end{figure}

      Figure~\ref{MnHXPES} shows Mn 2$p$ core-level XPS/HX-PES spectra of LSMO. Panel (a) shows that the distinct shoulder at 639--640 eV in Mn 2$p_{3/2}$ peak in HX-PES spectra is still observable in XPS spectra of $x=0.0$ and 0.4 although the intensity is smaller due to lower bulk sensitivity.\cite{Horiba04,Hishida12}
	Panel (b) shows the HX-PES spectra in the Mn 2$p_{3/2}$ region. The spectra of $x=0.0$ and 0.4 show very good agreement with those of thin films.\cite{Horiba04} In addition to this agreement, one can find the shoulder structure S even at $x=0.8$, for the first time. The systematic decrease in intensity and shift in energy of the shoulder S undoubtedly confirms that this feature corresponds to the Mn$^{3+}$ well-screened state. This shoulder is too small to observe in the XPS spectrum of $x=0.8$, and can be identified only by multiple-peak fitting.\cite{Hishida12}

\begin{figure}[t]
	\begin{center}
 	\includegraphics[width=80mm,keepaspectratio]{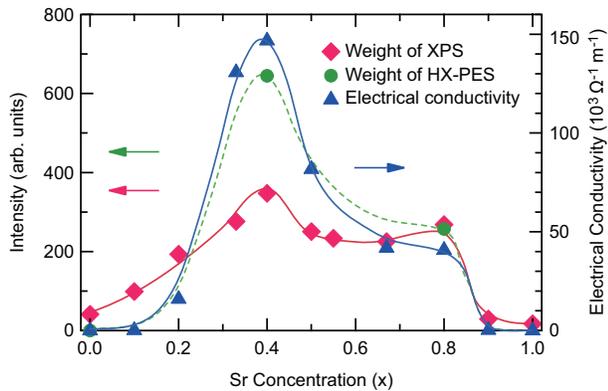}
  \caption{Comparison of $E_{\rm F}$ spectral weight of XPS (red diamonds) and electrical conductivity at the room temperature (blue triangles). $E_{\rm F}$ spectral weight of HX-PES (green circles) is also shown. The vertical scale of the HX-PES data is different from the XPS one because of different photoionization cross sections.}
\label{EC}
\end{center}
\end{figure}

	The $x$-dependence of the electrical conductivity at the room temperature (shown in Fig.~\ref{EC}) is as follows: At $x=0$, the system is insulating and shows an insulator-to-metal transition at $\sim$0.2 with increasing $x$. It is metallic between $\sim$0.2 and $\sim$0.8 with the maximum conductivity at 0.4 and becomes insulating again for a larger $x$.

	Hence, the intensity of S at $x=0.8$ is much smaller than at 0.0, while the electrical conductivity is opposite.
	This confirms our recent conclusion that this shoulder alone does not represent the metallic conductivity, but the sum of the spectral weight of the both Mn$^{3+}$ and Mn$^{4+}$ well-screened states does.\cite{Hishida12} This fact was not found so far because all previous studies were performed in the lower $x$-range.

\begin{figure}[b]
	\begin{center}
 	\includegraphics[width=80mm,keepaspectratio]{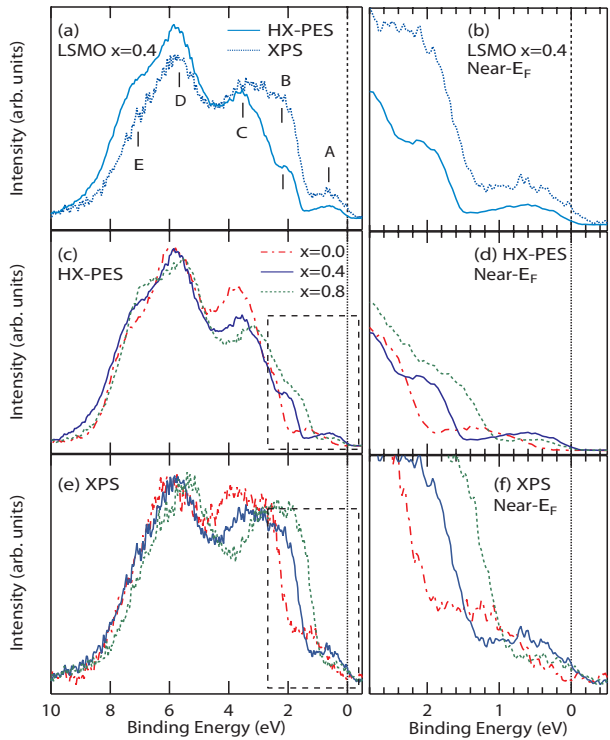}
  \caption{(a) Valence-band photoemission spectra of LSMO ($x=0.4$) by HX-PES and XPS. (b) and (c) Near-$E_{\rm F}$ spectra of LSMO by HX-PES and XPS.}
\label{VBHXPS}
\end{center}
\end{figure}


	Figure~\ref{VBHXPS} shows the valence-band HX-PES and XPS spectra.
The background intensity due to secondary electrons was subtracted using the Shirley's method and the valence-band intensity was corrected with respect to the Mn $2p_{3/2}$ core level intensity.\cite{Note1}
	In Fig.~\ref{VBHXPS}(a), one can identify five structures A--E, which are assigned to the Mn $3d\ e_g$ band (0.0--1.5 eV: A), the Mn $3d\ t_{2g}$ band ($\sim$2.0 eV: B), the O $2p$ nonbonding band (2.5--4.5 eV: C), the Mn $3d\ t_{2g}$--O $2p$ bonding band (5.0--6.5 eV: D), and the Mn $3d\ e_g$--O $2p$ bonding band (7.0--8.5 eV: E).\cite{Saitoh95, Horiba05}
	This interpretation is in agreement with the larger photoionization cross section of the $sp$ states compared to the $d$ states in HX-PES.\cite{Yeh85} In particular, the larger suppression of the Mn $3d\ t_{2g}$ band spectral weight shown in Fig.~\ref{VBHXPS}~(b) is a manifestation of smaller hybridization with the O $2p$ states than the $e_g$ band.

	Figures~\ref{VBHXPS}(c) and \ref{VBHXPS}(e) compare the whole valence-band spectra of LSMO with different $x$ measured with different excitation energies ((c): HX-PES, (e): XPS), both of which show systematic changes with $x$. The entire O $2p$ band width increases with $x$, probably because the GdFeO$_3$-type distortion is reduced with $x$.\cite{Saitoh95}
	Enlarged views of these HX-PES and XPS spectra in a near-$E_{\rm F}$ region are shown in Figs.~\ref{VBHXPS}(d) and \ref{VBHXPS}(f), respectively.
	For both HX-PES and XPS, the $e_g$ band shifts towards lower binding energy and the spectral weight decreases monotonically with increasing $x$. As a consequence, the $E_{\rm F}$ spectral weight that should be responsible for electrical conductivity initially increases and then decreases with $x$.
	In order to track such small $E_{\rm F}$ spectral weight, HX-PES has the benefits of a deeper probing depth and larger signal-to-noise (S/N) ratio than XPS, while the smaller $d$-cross section and the competitive accessibility to HX-PES facilities balances out these benefits. For this reason, we track the $E_{\rm F}$ spectral weight of XPS for the full range of $x$.

\begin{figure}[t]
	\begin{center}
 	\includegraphics[width=80mm,keepaspectratio]{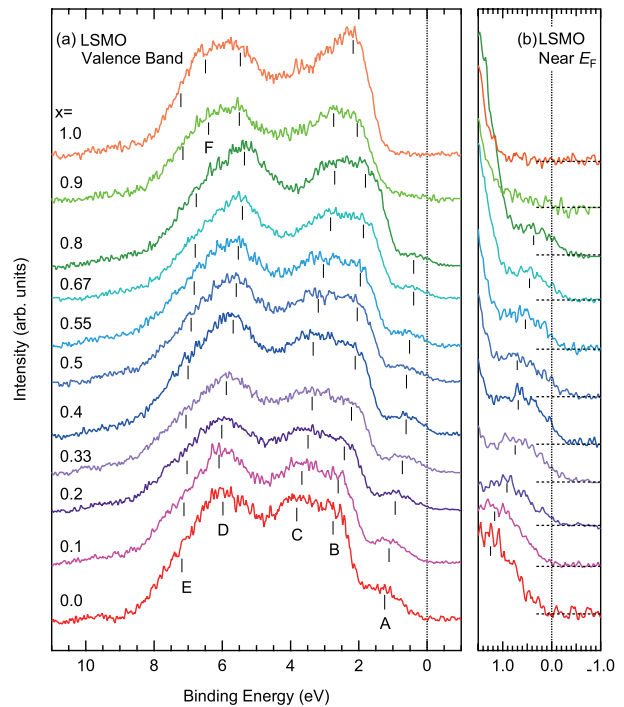}
  \caption{(a) Valence-band XPS spectra of LSMO with various Sr concentrations. (b) Near-$E_{\rm F}$ spectra of Panel (a).}
\label{VBXPS}
\end{center}
\end{figure}

	The complete valence-band XPS spectra of LSMO of the full range of $x$ are shown in Fig.~\ref{VBXPS}(a).
With increasing $x$, the features A--C rapidly move towards $E_{\rm F}$ first from $x=0$ to 0.33, but this shift slows down and almost pauses between $x=0.4$ and 0.55 and then resumes moving with a smaller amount than the first one between $x=0.55$ and 0.8.
	From $x=0.8$ to 0.9, the features B and C seem to move back away from $E_{\rm F}$ and a new feature F appears. This deviation from the systematics up to $x=0.8$ is probably due to a structural phase transition.\cite{Hishida12} However, the deviation does not affect the near-$E_{\rm F}$ electronic structure or electrical conductivity because the $e_g$ band fades away towards $x=1.0$.

	Within the rigid-band picture, the systematic energy shift (particularly of A) can be interpreted as a modulation of the $E_{\rm F}$ spectral weight (namely, the density of states at $E_{\rm F}$, $D(E_{\rm F})$) with $x$. This is because small energy shift $\Delta E$ due to a small amount of hole doping $\Delta x$ can be described as $\Delta x \propto D(E_{\rm F}) \Delta E$, hence giving the shift ratio $\Delta E/\Delta x$ proportional to $D(E_{\rm F})^{-1}$.\cite{Matsuno02}
	This simple interpretation indeed qualitatively explains the linked behavior of the $E_{\rm F}$ spectral weight and the energy shift shown in Fig.~\ref{VBXPS}(b), and qualitatively explains the trend of the electrical conductivity shown in Fig.~\ref{EC}.
	Nonetheless, the global evolution of the whole $e_g$ spectral weight never follows the rigid band behavior as already pointed out in the literature;\cite{Saitoh95} changing its line shape, the spectral weight decreases almost linearly from $x=0.0$ to 1.0 as seen in Fig.~\ref{VBXPS}(b).

	The $E_{\rm F}$ spectral weight of the XPS spectra in Fig.~\ref{VBXPS}(b) is plotted against $x$ in Fig.~\ref{EC} together with the electrical conductivity. The spectral weight at $E_{\rm F}$ was estimated by the integrated intensity in the window slightly larger than the experimental energy resolution centered at $E_{\rm F}$, $\pm 325$ meV.
	We see that the plotted $E_{\rm F}$ weight qualitatively or even semi-quantitatively reproduces the electrical conductivity variation with $x$. The spectral weight from the HX-PES measurement is also shown for comparison, which shows improved agreement with the conductivity. This difference is probably caused by a combination of larger surface sensitivity of XPS than HX-PES and reduced ferromagnetism and metallicity at the surface.\cite{Park98}

	The observed qualitative agreement between the $E_{\rm F}$ weight and the conductivity is rather surprising because the integration window (650 meV) is much larger than the reported quasiparticle line width ($\sim$50 meV),\cite{Mannella05,Sun06} and we should not be able to track the $x$-dependent change of the quasiparticle spectral weight.
	This situation reminds us of the case of the layered manganites, namely, the temperature-dependence of the near-$E_{\rm F}$ incoherent spectral weight (spreading over hundreds of meV below the quasiparticle) which mirrors that of electrical conductivity\cite{Saitoh00} and the quasiparticle spectral weight.\cite{Mannella07}
	In the layered manganites, the near-$E_{\rm F}$ incoherent spectral weight is transferred to higher binding energy with increasing temperature, which causes the above agreement.
	The same type of spectral-weight transfer with $x$ modulation can be expected in the present three-dimensional system, as observed in La$_{1-x}$Sr$_x$TiO$_3$.\cite{Yoshida02}
	As a result, the XPS near-$E_{\rm F}$ spectral weight integrated over a few hundreds meV can be a rough measure of the total quasiparticle spectral weight in spite of a limited energy resolution of XPS.


\section{Conclusions}

	We have performed x-ray and hard-x-ray photoemission spectroscopy study on the magnetoresistive oxide La$_{1-x}$Sr$_x$MnO$_3$ in order to determine a relation between the valence/core-level photoemission spectral weight and the electrical conductivity as a function of $x$.
	We observed a small Mn$^{3+}$ well-screened final state for the $x=0.8$ sample for the first time, which confirms this well-screened state alone cannot be a measure of metallicity.\cite{Hishida12}
	We found that the near-$E_{\rm{F}}$ XPS spectral weight can be a measure of the variation of electrical conductivity with $x$ in spite of a much lower energy resolution compared with the energy scale of the quasiparticle peak. This is because the change of the quasiparticle spectral weight and the spectral weight transfer of the incoherent part appear concurrently.

\begin{acknowledgments}
We would like to thank H.\ Kozuka for providing us with LSMO samples used in this work. T.S.\ thanks Y.\ Takamura for helpful discussions. The synchrotron radiation experiments at SPring-8 were performed under the approval of the Japan Synchrotron Radiation Research Institute (Proposal Numbers 2011A1420 and 2011B1710).
\end{acknowledgments}

\onecolumngrid
\clearpage
\begin{center}
\large{\textbf{Supplemental Material}}
\end{center}

\begin{figure}[h]
	\begin{center}
 	\includegraphics[width=100mm,keepaspectratio]{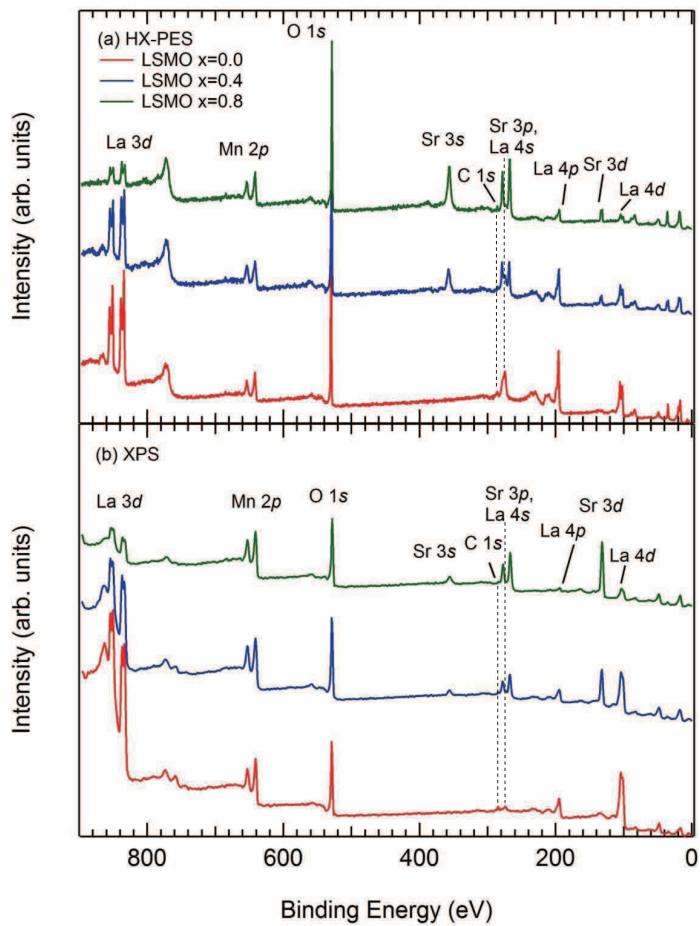}
  \caption{(a) HX-PES and (b) XPS survey spectra of LSMO ($x=0$, 0.4, and 0.8).}
\label{SFig}
\end{center}
\end{figure}

\large{Supplementary Figure \ref{SFig} shows the survey scans for (a) HX-PES and (b) XPS measurements.
The right (left) vertical broken lines at about 280 eV denote the La $4s$ (C $1s$) peaks.
It is obvious that the C $1s$ intensity is negligibly small in both HX-PES and XPS survey spectra,
demonstrating that the quality of the measured sample surfaces was good enough for our purpose.}


\begin{thebibliography}{99}
\bibitem{vonHelmolt93}R. von Helmolt, J. Wecker, B. Holzapfel, L. Schultz, and K. Samwer,  Phys. Rev. Lett. {\bf 71}, 2331 (1993).
\bibitem{Tokura94}Y. Tokura, A. Urushibara, Y. Moritomo, T. Arima, A. Asamitsu, G. Kido, and N. Furukawa, J. Phys. Soc. Jpn. {\bf 63}, 3931 (1994).
\bibitem{Moritomo96}Y. Moritomo, A. Asamitsu, H. Kuwahara, and Y. Tokura, Nature \textbf{380}, 141 (1996).
\bibitem{Cohn97}J. L. Cohn, J. J. Neumeier, C. P. Popviciu, K. J. McClellan, and Th. Leventouri, Phys. Rev. B {\bf 56}, R8495 (1997).
\bibitem{Averitt01}R. D. Averuttm, A. I. Lobad, C. Kwon, S. A. Trugman, V. K. Thorsmolle, and A. J. Taylor, Phys. Rev. Lett. {\bf 87}, 017401 (2001).

\bibitem{Chainani93}A. Chainani, M. Mathew, and D. D. Sarma, Phys. Rev. B {\bf 47}, 15397 (1993).
\bibitem{Saitoh95}T. Saitoh, A. E. Bocquet, T. Mizokawa, H. Namatame, A. Fujimori, M. Abbate, Y. Takeda, and M. Takano, Phys. Rev. B {\bf 51}, 13942 (1995).
\bibitem{Zener51}C. Zener, Phys. Rev. \textbf{82}, 403 (1951).
\bibitem{Millis95}A. J. Millis, P. B. Littlewood, and B. I. Shraiman, Phys. Rev. Lett. \textbf{74}, 5144 (1995).
\bibitem{Horiba04}K. Horiba, M. Taguchi, A. Chainani, Y. Takata, E. Ikenaga, D. Miwa, Y. Nishino, K. Tamasaku, M. Awaji, A. Takeuchi, M. Yabashi, H. Namatame, M. Taniguchi, H. Kumigashira, M. Oshima, M. Lippmaa, M. Kawasaki, H. Koinuma, K. Kobayashi, T. Ishikawa, and S. Shin, Phys. Rev. Lett. {\bf 93}, 236401 (2004).
\bibitem{Ueda09}S. Ueda, H. Tanaka, E. Ikenaga, J. J. Kim, T. Ishikawa, T. Kawai, and K. Kobayashi, Phys. Rev. B \textbf{80}, 092402 (2009).
\bibitem{Hishida12}T. Hishida, K. Ohbayashi, and T. Saitoh, J. Appl. Phys. {\bf113}, 043710 (2013).
\bibitem{Urushibara95}A. Urushibara, Y. Moritomo, T. Arima, A. Asamitsu, G. Kido, and Y. Tokura, Phys. Rev. B {\bf 51}, 14103 (1995).
\bibitem{Sarma96}D. D. Sarma, N. Shanthi, S. R. Krishnakumar, T. Saitoh, T. Mizokawa, A. Sekiyama, K. Kobayashi, A. Fujimori, W. Weschke, R. Meier, G. Kaindl, Y. Takeda, and M. Takano, Phys. Rev. B \textbf{53}, 6873 (1996).
\bibitem{Park96}J.-H. Park, C. T. Chen, S.-W. Cheong, W. Bao, G. Meigs, V. Chakarian, and Y. U. Idzerda, Phys. Rev. Lett. {\bf 76}, 4215 (1996).
\bibitem{Yoshida02}T. Yoshida, A. Ino, T. Mizokawa, A. Fujimori, Y. Taguchi, T. Katsufuji, and Y. Tokura, Europhys. Lett. {\bf 59}, 258 (2002).
\bibitem{Mannella05}N. Mannella, W. L. Yang, X. J. Zhou, H. Zheng, J. F. Mitchell, J. Zaanen, T. P. Devereaux, N. Nagaosa, Z. Hussain, and Z.-X. Shen, Nature {\bf 438}, 474 (2005).

\bibitem{Sun06}Z. Sun, Y.-D. Chuang, A. V. Fedorov, J. F. Douglas, D. Reznik, F. Weber, N. Aliouane, D. N. Argyriou, H. Zheng, J. F. Mitchell, T. Kimura, Y. Tokura, A. Revcolevschi, and D. S. Dessau, Phys. Rev. Lett. {\bf 97}, 056401 (2006).
%
\bibitem{Saitoh00}T. Saitoh, D. S. Dessau, Y. Moritomo, T. Kimura, Y. Tokura, and N. Hamada, Phys. Rev. B {\bf 62}, 1039 (2000).
%
\bibitem{Mannella07}N. Mannella, W. L. Yang, K. Tanaka, X. J. Zhou, H. Zheng, J. F. Mitchell, J. Zaanen, T. P. Devereaux, N. Nagaosa, Z. Hussain, and Z.-X. Shen, Phys. Rev. B \textbf{76}, 233102 (2007).

\bibitem{Kobata10}M. Kobata, I. Pis, H. Iwai, H. Yamazui, H. Takahashi, M. Suzuki, H. Matsuda, H. Daimon, and K Kobayashi, Anal. Sci. {\bf 26}(2), 227 (2010).

\bibitem{Bindu11}R. Bindu, G. Adhikary, N. Sahadev, N. P. Lalla, and K. Maiti, Phys. Rev. B {\bf 84}, 052407 (2011).

\bibitem{Suppl}See supplementary material for the quality of the sample surface monitored by survey scans.

\bibitem{Note1}The spectral weight of the Mn 2$p_{3/2}$ core-level spectra from 636 eV to 648 eV after background subtraction with the Shirley's method was used for the intensity normalization of the valence band.

\bibitem{Horiba05}K. Horiba, A. Chikamatsu, H. Kumiashira, M. Oshima, N. Nakagawa, M. Lippmaa, K. Ono, M. Kawasaki and H. Koinuma, Phys. Rev. B {\bf71}, 155420 (2005).

\bibitem{Yeh85}J. J. Yeh and I. Lindau, At. Data Nucl. Data Tables {\bf32}, 1 (1985).

\bibitem{Matsuno02}A similar formula was given by J. Matsuno, A. Fujimori, Y. Takeda, and M. Takano, Europhys. Lett. {\bf 59}, 252 (2002).


\bibitem{Park98}J.-H. Park, E. Vescovo, H.-J. Kim, C. Kwon, R. Ramesh, and T. Venkatesan, Phys. Rev. Lett. {\bf 81}, 1953 (1998).



\end{thebibliography}
\end{document}